\documentclass[11pt]{article}
\usepackage[utf8]{inputenc}
\usepackage{graphicx}
\usepackage{ar}
\usepackage{pgfplots}
\usepackage[numbers]{natbib} 
\usepackage[modulo]{lineno}
\usepackage{gacaps}
\usepackage[hidelinks]{hyperref}

\usepackage[version=4]{mhchem}
\usepackage{siunitx}
\usepackage{longtable,tabularx}
\usepackage[top=2cm, bottom=2cm, left=1.1in, right=1.1in]{geometry}

\pgfplotsset{width=7cm,compat=1.14}

\definecolor{red}{rgb}{   1,  0,  0}
\definecolor{blue}{rgb}{  0,  0,  1}
\definecolor{green}{rgb}{ 0, .5,  0}
\definecolor{green1}{rgb}{0, .7,  0}
\definecolor{black}{rgb}{ 0,  0,  0}
\definecolor{grey}{rgb}{ .5, .5, .5}

\definecolor{matlabdef1}{rgb}{ 0, .447, .741}
\definecolor{matlabdef2}{rgb}{ 0.85, .325, .098}
\definecolor{matlabdef3}{rgb}{ 0.929, .694, .125}
\definecolor{matlabdef4}{rgb}{ 0.494, .184, .556}
\definecolor{matlabdef5}{rgb}{ 0.466, .674, .188}

\newcommand{\lyy}[2]{\lineSymbol[#1]{none}{#2}{.8ex}{1pt}{#2}{#2}}

\title{Numerical simulation of flow over flapping wings in tandem: wingspan effects} 

\author{R. Jurado$^1$, G. Arranz$^2$, O. Flores$^{1,*}$, and M. Garc\'{i}a-Villalba$^1$ \\[2ex]
$^1$ \small{Bioengineering and Aerospace Eng. Dept., Universidad Carlos III de Madrid, Spain} \\[1ex] 
$^2$ \small{Dept. of Aeronautics and Astronautics, Massachusetts Institute of Technology, Cambridge, USA} \\[1ex]
\small $^*$Corresponding author: oflores@ing.uc3m.es
 }

\begin{document}

\maketitle

\begin{abstract}

We report direct numerical simulations of a pair of wings in horizontal tandem configuration, to analyze the effect of their aspect ratio on the flow and the aerodynamic performance of the system. The wings are immersed in a uniform free-stream at Reynolds number $Re=1000$, and they undergo heaving and pitching oscillation with Strouhal number $St=0.7$. The aspect ratios of forewing and hindwing vary between 2 and 4. The aerodynamic performance of the system is dictated by the interaction between the trailing edge vortex (TEV) shed by the forewing and the induced leading edge vortex formed on the hindwing. The aerodynamic performance of the forewing is similar to that of an isolated wing irrespective of the aspect ratio of the hindwing, with a small modulating effect produced by the forewing-hindwing interactions. On the other hand, the aerodynamic performance of the hindwing is clearly affected by the interaction with the forewing’s TEV. Tandem configurations with a larger aspect ratio on the forewing than on the hindwing result in a quasi-two-dimensional flow structure on the latter.
This yields an 8\% increase in the time-averaged thrust coefficient of the hindwing, with no change in its propulsive efficiency.
\end{abstract}

\section{Introduction}

The capabilities of flapping-wings animals like birds, bats and insects, have attracted for years the attention of the scientific community.
From an engineering perspective, one of the main drivers of this attention is the desire to develop bio-inspired devices that mimic animal fliers, in the hope of reproducing their flying capabilities \citep{croon2009,tsai2009,keennon2012}.
However, despite the recent improvements in the understanding of unsteady aerodynamics mechanisms \citep{shyy2013}, achieving a similar performance to that of actual animal flight is still a challenge.
The multiple interplaying flow mechanisms involved in this type of unsteady aerodynamics make this a complex problem \citep{sane2003,platzer2008flapping,haider2020recent}.
Among the flying animals, dragonflies might be specially interesting for the developing of bio-inspired devices, since the independent control of the motion of each pair of wings allows these insects to adopt many different flying configurations.
The flight of dragonflies has been indeed studied for a long time by many authors \citep{alexander1984,wakeling1997,thomas2004,hefler2020}.
The key unsteady aerodynamic mechanism exploited by dragonflies is the wing-wing flow interaction between hind- and forewings, analogous to the wake capture mechanism observed in insects with a single pair of wings \citep{sane2003,diaz2021,han2019}. 
Different studies have shown that dragonflies take advantage of these wake interactions to enhance their maneuverability capabilities \citep{wang2007effect,usherwood2008phasing,lehmann2008wings}, being even able to sustain backward flight \citep{bode2018flying}.

Although some authors have analyzed realistic dragonfly-like wing and kinematic models \citep{isogai2004,kamisawa2008optimum,li2017wing,nagai2019}, we are still not capable to design efficiently micro air vehicles with two pairs of flapping wings. 
To overcome this limitation, a better understanding of the fundamentals of unsteady aerodynamics of flapping wings in tandem is still needed.
To explore the underlying effects of the flow interaction among multiple wings arrangements, in general, the complexity of the model has to be sacrificed. 
The need for simplification becomes obvious when the problem of optimizing the kinematics of flapping wings in tandem configuration is considered. Even in the simplest case (forward flight, sinusoidal heave and pitch motion for the wings), the parametric space associated to the geometry (i.e., planform shape, aspect ratio, separation between the wings) and kinematics (frequency, amplitude and relative offsets between the heaving and pitching motions of each wing) of the pair of wings is huge \citep{platzer2008flapping}. 
Indeed, optimization studies using simulations 
\citep{ortega2016analysis, Ortega2019, kamisawa2008optimum}
or experiments 
\citep{huera2018propulsive}
severely restrict the parametric space, performing the optimization over a reduced number of variables. 
Note that this is necessary even if the Reynolds number of these flyers is small ($10^1 < Re < 10^4$ \citep{wang2005dissecting}), which should be beneficial from the point of view of the computational cost associated to the numerical simulations.

There are several works in the literature which analyze the benefits of horizontal tandem configurations in forward flight, compared with its isolated wing counterpart, both for idealized two-dimensional (2D) 
\citep{rival2010,rival2011b,lua2016aerodynamics,muscutt2017performance,alaminos2020aerodynamics} and three-dimensional (3D) configurations \citep{broering2015numerical,kurt2018flow}, using relatively simple geometries (i.e., flat plates, rectangular or elliptical wings) and kinematics (sinusoidal heaving and/or pitching motions). 
In idealized configurations the interaction of the hindwing with the vortical structures shed by the forewing typically results in an enhancement of the aerodynamic forces compared to the corresponding isolated flapping wings. 
Akhtar et al \citep{akhtar2007hydrodynamics} used 2D simulations of a pair of rigid foils in tandem arrangement to show that the trailing edge vortex (TEV) shed by the forewing can increase the effective angle of attack of the hindwing. This results in the generation of a leading edge vortex on the hindwing (also referred to as induced leading edge vortex, iLEV, in other studies) that increases thrust, although this increment is very sensitive to the phase of both wings. 
Whether this interaction results in positive or negative effects on the aerodynamic performance of the system depends on many factors.
For instance, Broering and Lian \citep{broering2013investigation} 
showed that in-phase flapping leads to high aerodynamic force production, whereas counter-phase flapping results in high power efficiency, both in 2D and 3D. 
Ortega-Casanova and Fern\'andez-Feria \citep{Ortega2019} used a large database of 2D simulations of two flat plates undergoing a pure heaving motion  to analyze the effect of the frequency and phase shift between the two plates on the propulsive efficiency.
They found that, for their configuration, the most efficient heaving motion was counterstroking (i.e., a phase shift of $180^{\circ}$).
Similarly the spacing between the two wings can be varied to improve the propulsive efficiency or the thrust generation since these two parameters control the timing between the formation of the trailing edge vortex and its arrival to the hindwing \citep{maybury2004fluid,boschitsch2014propulsive, lua2016aerodynamics, muscutt2017performance}.

One of the major drawbacks of studying 2D configurations consists on the uncertainty associated to the translation of the 2D results into a 3D configuration. 
In other words, how the aerodynamic performance of airfoils (2D) does compare to that of wings (3D). 
For instance,  Broering and Lian \citep{broering2013investigation} found that the vortical  structures formed in the three-dimensional configurations were less intense than their two-dimensional counterparts.
They related this observation to the wing tip vortices, that partially suppress the formation of the iLEV near the wing tips, weakening the iLEV at the midspan too. 
More recently, Arranz et al \citep{Arranz2020} used  optimal kinematics for a 2D configuration \citep{Ortega2019} to compare the aerodynamic performance of the 2D case with two different 3D cases, with aspect ratios 2 and 4 (same aspect ratio for fore- and hindwing). 
They found that the three-dimensional effects result in a reduction of the aerodynamic forces with respect to the two-dimensional configuration, consistently with previous studies \citep{broering2013investigation}.
They also found that the flow around the hindwing was affected by the wingtip vortices shed from the forewing in a similar way for wings of aspect ratio 4 and 2. 
In both cases, the  flow near the hindwing-tips was found to be highly three-dimensional, possibly having a detrimental effect on the aerodynamic performance of the hindwing.

Among all the possible parameters involved in the flow over multiple oscillating wings at low Reynolds numbers, the spanwise effects of the wake shed by a leading wing on a following wing seems to be somewhat unexplored.
%
%In \cite{garmann2014unsteady,garmann2015interaction}, 
Garmann and coworkers \citep{garmann2014unsteady,garmann2015interaction} 
have studied the effect on the aerodynamic forces of an analytical stream-wise oriented vortex impinging at different offset locations from the tip of a downstream finite wing.
They have found that the lift-to-drag ratio have a maximum when the impinging vortex is aligned with the tip of the wing and decays when the vortex is offset either inboard or outboard the tip.
However, their analysis is more focused on the steady aerodynamics that appear in formation flying of fixed-wing aircrafts, rather than in the unsteady impinging of vortices in a flapping wing.
Recently, Chen et al \citep{chen2018experiments} have characterized experimentally the flow structures generated by the interaction of the wing tip vortices of two wings with different vertical and spanwise offsets for the wing tips at $Re = 5 \cdot 10^3$.
Similarly, McKenna and coworkers \citep{mckenna2018interaction,mckenna2017structure} have investigated the flow patterns formed by an impinging wing tip vortex upon the wing tip of an oscillating wing.
However, these studies mainly focus on the flow structures, providing little information on the aerodynamic performance of the wings. 

The present study has been designed to address the following questions: How does the wing to wing aspect ratio affect the generation of forces on each wing of an oscillating tandem configuration?
 Is it possible to reduce the influence of the wing tip vortices of the forewing on the aerodynamic performance of the hindwing and its spanwise loading distribution? 
Thus, to answer these questions we have developed and analyzed a database of numerical simulations of heaving and pitching rigid wings in forward flight in horizontal tandem configuration, varying independently the aspect ratio of forewing and hindwing.
The paper is organized as follows; the geometric and kinematic parameters, numerical method and definition of the aerodynamics coefficients are described in Section \ref{method}. In Section \ref{results} the results are discussed in terms of 
 flow visualizations, global aerodynamic force coefficients, sectional force coefficients and power requirements. Finally, the conclusions of the study are summarized in Section \ref{conclusions}.

\section{Methodology}
\label{method}
\subsection{Problem description}

\begin{figure}
   \centering
    \includegraphics[width=0.65\textwidth]{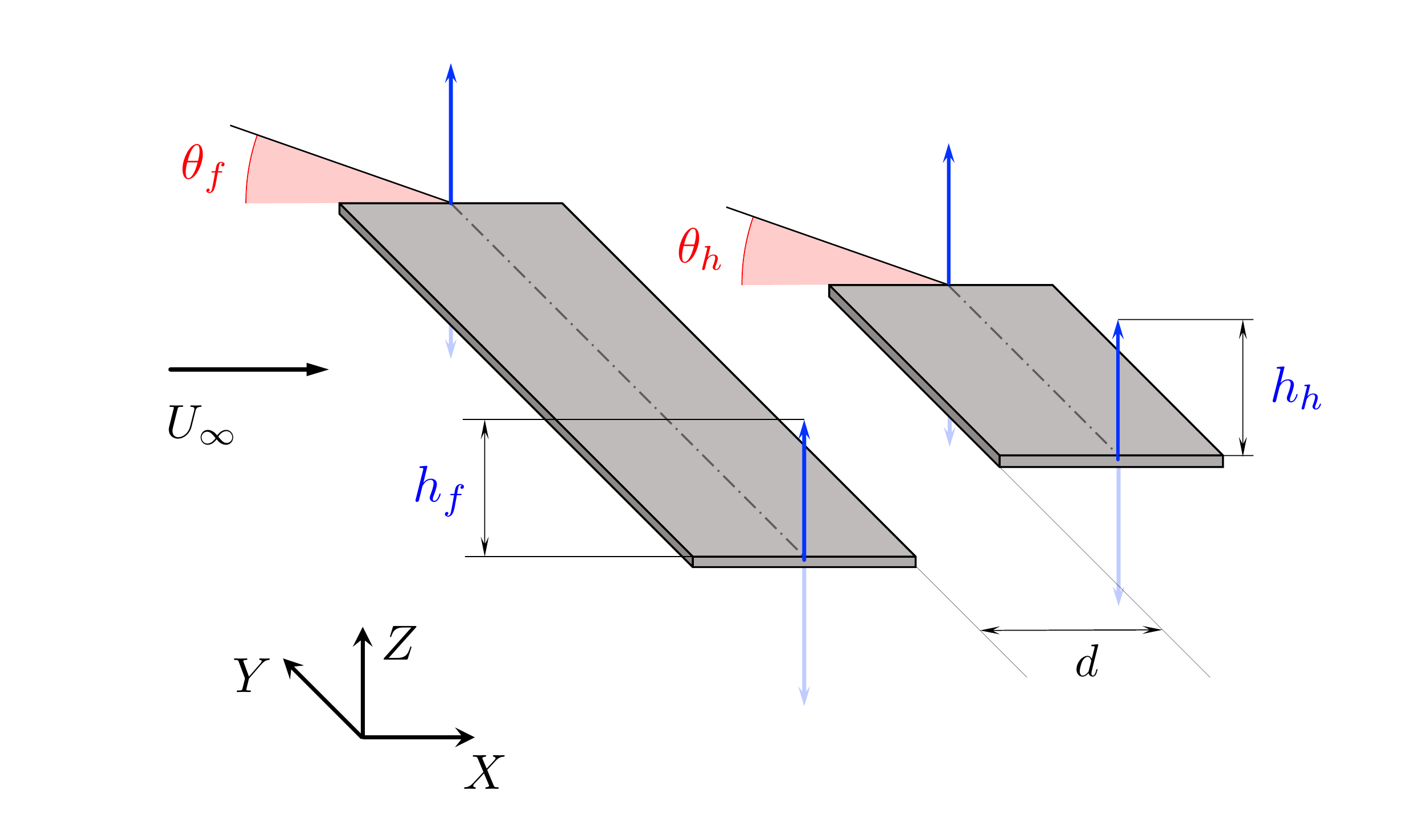}
   \caption{Sketch of the wings in tandem arrangement.}
   \label{fig:fig1}
\end{figure}

Two finite wings in a horizontal tandem configuration immersed in a uniform free-stream of magnitude, $U_{\infty}$, are considered.
The two wings correspond to flat plates with rectangular planform with chord, $c$, and thickness $e = c/96$.
The planform area of the forewing (hindwing) is $S_f$ ($S_h$). 
The distance between the trailing edge of the forewing and the leading edge of the hindwing when they lay onto a horizontal plane is $d/c = 0.5$ (see Figure \ref{fig:fig1}). 
The Reynolds number based on the chord and the free-stream velocity is $Re = U_{\infty} c / \nu = 1000$ for all the cases.

%%%%%%%%%%%%%%%%%%%%%%%%%%%%%%%%%%%%%%%%%%%%%%%%%%%%%%%%%%%%%%%%%%%%%%%%%%%%%%%
%%%%%%%%%%%%%%%%%%%%%%%%%%%%%%%%%%%%%%%%%%%%%%%%%%%%%%%%%%%%%%%%%%%%%%%%%%%%%%%

The kinematics of the system is based on an optimal, two-dimensional configuration \citep{Ortega2019}. 
The motion of both wings is a combination of heaving and pitching about the mid-chord. The motion law for the pitching angle, $\theta_0$, and heaving amplitude, $h_0$, are
\begin{align}
    h_i (t) &= h_0 \cos( 2 \pi f t + \varphi_{h,i}),
    \label{eq:heaving}\\
    \theta_i (t)& = \theta_0 \cos( 2 \pi f t + \varphi_{\theta,i}),
    \label{eq:pitching}
\end{align}
where the subscript $i$ indicates the wing ($i = f$ forewing, $i = h$ hindwing) as shown in Figure \ref{fig:fig1}.
The sinusoidal motion of the wings is defined by the heaving, $h_0$, and pitching, $\theta_0$, amplitudes, the frequency, $f$, and the heaving and pitching phase shifts, $\varphi_{h,i}$, and $\varphi_{\theta,i}$, respectively. 
The frequency and heaving amplitude are fixed by the Strouhal number based on the chord length, $St_c = fc/U_{\infty}$, and the heaving amplitude, $St_a = h_0f /U_{\infty}$. All the parameters that define the motion are gathered in Table \ref{tab:numbers}. 
The mean pitch angle is zero, therefore the motion is symmetric with respect to the horizontal plane. Note that, according to the values of $\varphi_{h,i}$ and $\varphi_{\theta,i}$ in Table \ref{tab:numbers}, pitching is advanced $3\pi/4$ with respect to heaving for both wings. 
Consequently, the motion of both wings is identical with a phase shift equal to $\pi$.

In this study, we consider five combinations of aspect ratios for the forewing ($\AR_f$) and hindwing ($\AR_h$), all satisfying $\AR_f \ge \AR_h$.  
Cases with identical fore- and hindwings (i.e., $\AR_f=\AR_h=2$ and $\AR_f=\AR_h=4$) are obtained from a previous study \citep{Arranz2020}. 
Three new cases with $\AR_f > \AR_h$ are simulated: 
$\AR_f = 4$ with $\AR_h=3$ and 2, and $\AR_f=3$ with $\AR_h=2$. 
These five cases are labeled using the acronym ``AR'' followed by two digits denoting the aspect ratio of the forewing and hindwing respectively (i.e., AR32 for $\AR_f=3$ and $\AR_h=2$).
Two simulations of isolated wings, with aspect ratios 2 and 4, have also been performed to complete the discussion of the results.

\begin{table}
\begin{center}
\begin{tabular}{cccccc}
             
              & $\varphi_{h,i}$ & $\varphi_{\theta,i}$ & $\theta_0$  & $St_c$ & $St_a$   \\
             \hline
             forewing & $0$ & $3 \pi / 4$ & $25^\circ $ & $0.7$ & $0.17$   \\
             %\hline
             hindwing & $\pi$ & $- \pi / 4$ & $25^\circ $ & $0.7$ & $0.17$

\end{tabular}
\end{center}
	\caption{Parameters of the motion law for both wings.}
	\label{tab:numbers}
\end{table}

\subsection{Computational set-up}

The flow around the pair of flapping wings desccribed in the previous section is simulated with TUCAN, an in-house, parallel solver for the incompressible Naiver-Stokes equations. 
TUCAN uses second-order finite differences for the spatial discretization in a staggered grid and a 3-stage, low-storage Runge–Kutta scheme for time integration. 
The time step is always selected so that the Courant–Friedrichs–Lewy number is smaller than 0.2. 
%
%The presence of the bodies is modeled by means of a direct forcing formulation of the immersed %boundary method \citep{Uhlmann2005}. 
%
%{\color{blue}
The wings are modeled using an immersed boundary method (IBM) \cite{mittal2005}.
This is a technique in which a body force is added to the momentum equation in order to 
fulfill the no-slip condition at the boundary of the object. 
There is a large number of variations of the IBM as reviewed, for example, by 
Sotiropoulos and Yang \cite{sotiropoulos2014}.
In this work we employ the direct forcing formulation proposed by Uhlmann \cite{Uhlmann2005}.
%}
%
TUCAN has been successfully used for the simulation of aerodynamic flows, both in two- \citep{Moriche2017,martinez2020analysis} and three-dimensions \citep{Moriche2016,Gonzalo2018,Arranz2018b,Arranz2018a,moriche2021,moriche2020c,arranz2022}.

\begin{figure}
\centering
    %\begin{subfigure}{0.6\textwidth}
        %\includegraphics[trim=75 25 50 25,clip,width=0.6\textwidth]{fig2.eps}
        \includegraphics[trim=75 25 50 25,clip,width=0.6\textwidth]{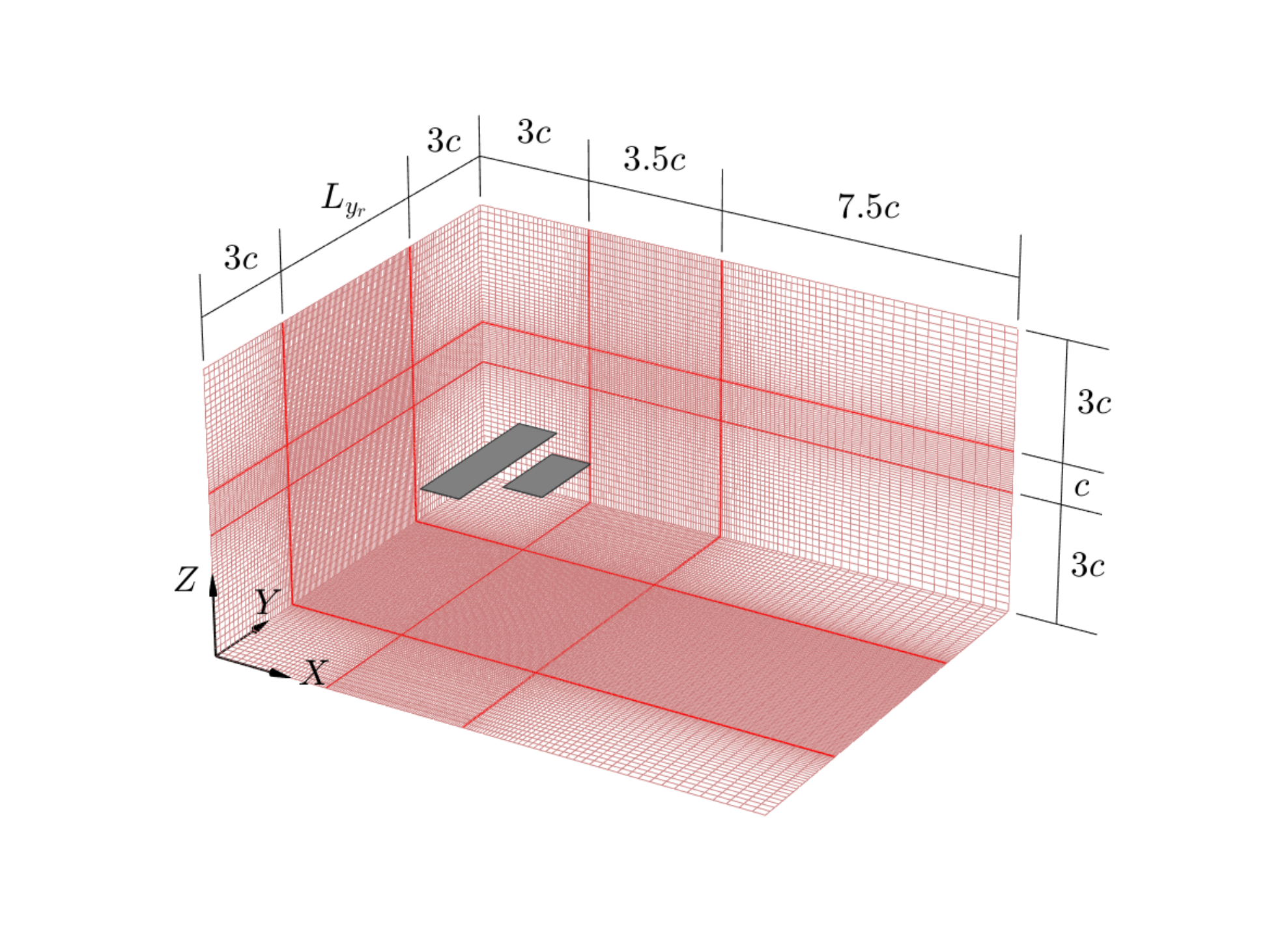}

    \caption{Sketch of the computational domain}
    \label{fig:domain}
\end{figure}

The computational domain is a rectangular prism, shown in Figure \ref{fig:domain}.
The wings are centred in a refined region ($3.5c \times L_{y_r} \times 1c$) with an uniform grid spacing in all directions, $\Delta r = c/96$.
Outside this region, a constant stretching of $1\%$ is applied to the grid in all directions.
The length, $L_{y_r}$, of the refined region is adapted to the aspect ratio of the forewing, $\AR_{f}$, of the corresponding case as $L_{y_r}/c = 1 + \AR_{f}$ in order to keep the same distance between the wing-tips and the boundaries for all cases.
As a result, the computational domain is discretized in $\sim 1.41 \times 10^{8}$ grid points for the AR22 case and $\sim 1.90 \times 10^{8}$ for the rest of the cases.

Regarding the boundary conditions, a uniform free-stream velocity, $U_{\infty}$, is imposed at the inflow plane ($X=0$). 
A convective boundary condition is imposed at the outflow plane ($X = 14c$).
Free-slip boundary conditions are imposed at the lateral boundaries.
 
The immersed boundary method in TUCAN requires the specification of a Lagrangian mesh for the wings.
Since the wings are flat plates, two flat surfaces are employed to discretize each wing.
A uniform grid is used for each surface, with a grid spacing $\Delta r = c/96$ in both spanwinse and chordwise directions.
%

%%%%%%%%%%%%%%%%%%%%%%%%%%%%%%%%%%%%%%%%%%%%%%%%%%%%%%%%%%%%%%%%%%%%%%%%%%%%%%%
% FIGURE COMPUTATIONAL DOMAIN SKETCH AND TYPICAL FLOW VISUALIZATION
%%%%%%%%%%%%%%%%%%%%%%%%%%%%%%%%%%%%%%%%%%%%%%%%%%%%%%%%%%%%%%%%%%%%%%%%%%%%%%%

The grid resolution was chosen after performing a grid sensitivity analysis in a 2D configuration for $\Delta r = c / 48$, $\Delta r = c / 96$ and $\Delta r = c / 192$. 
The variation of the propulsive efficiency when comparing $\Delta r = c / 96$ and $\Delta r = c / 192$ were about $3\%$.
Therefore the grid spacing $\Delta r = c / 96$ was selected, since it showed the best ratio accuracy/computational cost. For more details about the grid sensitivity analysis, see \citep{Arranz2020}. 
All simulations are started using a low resolution grid ($\Delta r = c/56$ in the refined region) during 3 cycles, then the flow field is interpolated on the finer grid and the simulations are restarted and run until convergence is achieved. This entails about $3 - 4$ additional cycles.
For the present cases, convergence means that periodic flow conditions are obtained for all cases, with the same period of oscillation of the forcing motion.
Consequently, the aerodynamic forces and the flow in the region of interest are periodic,
and the discussion of the results is based on the last computed cycle without any loss of generality.

\subsection{Definition of the aerodynamic coefficients}

The aerodynamic force coefficients are defined as
\begin{equation}
    C_{k,i} = \frac{2 \mathbf{F}_i \cdot \mathbf{e}_k}{\rho U_\infty^2 S_i} 
    \label{eq:3dcoeff}
\end{equation}
where $\mathbf{F}_i$ is the total aerodynamic force on the i-wing, $\mathbf{e}_k$ is the unitary vector parallel to the k-axis and $\rho$ is the fluid density. On the other hand, we also analyze sectional forces at a given spanwise position. To that end, we define the sectional force coefficients as
\begin{equation}
    c_{k,i} ( y) = \frac{2 \mathbf{f}_i ( y ) \cdot \mathbf{e}_k}{\rho U_\infty^2 c}
    \label{eq:2dcoeff}
\end{equation}
where $\mathbf{f}_i ( y )$ is the sectional force at the spanwise position $y$ of the i-wing. 

Finally, the performance of the wings is assessed by means of the individual propulsive efficiency of each wing, computed as:
    \begin{equation}
        \eta_{p,i} = \frac{\overline{C}_{T,i}}{\overline{P}_i},
	\label{eq:effi}
    \end{equation}
where $\overline{C}_{T,i}$, is the average thrust coefficient (computed as the average of $-C_{x,i}$ over a cycle), and $\overline{P}_{i}$ is the averaged non-dimensional input power of the i-wing over a cycle. The instantaneous power is computed as:
\begin{equation}
    P_i (t) = \max\left(-C_{z,i} \frac{\dot{h}_i (t)}{U_\infty},0\right) + \max\left(-\frac{2 {M}_{y,i} \dot{\theta_i}(t)}{\rho U_\infty^3 S_i},0\right) 
\label{eq:powerinput}
\end{equation}
where $M_{y,i}$ is the spanwise component of the aerodynamic moment with respect to the pivoting axis of the $i$-wing.
The first term in Eq. \eqref{eq:powerinput} represents the power requirements associated to the heaving motion of the wing, while the second term represents the power requirements associated to the pitching motion of the wing. 
Finally, note that this definition of the instantaneous power models a system in which the actuators can not extract energy from the fluid \citep{berman2007,vejdani2018}.

\section{Results}
\label{results}

\subsection{Flow structures} \label{sec:flow_struct}

\begin{figure}
   \centering
	\includegraphics[width=1\textwidth]{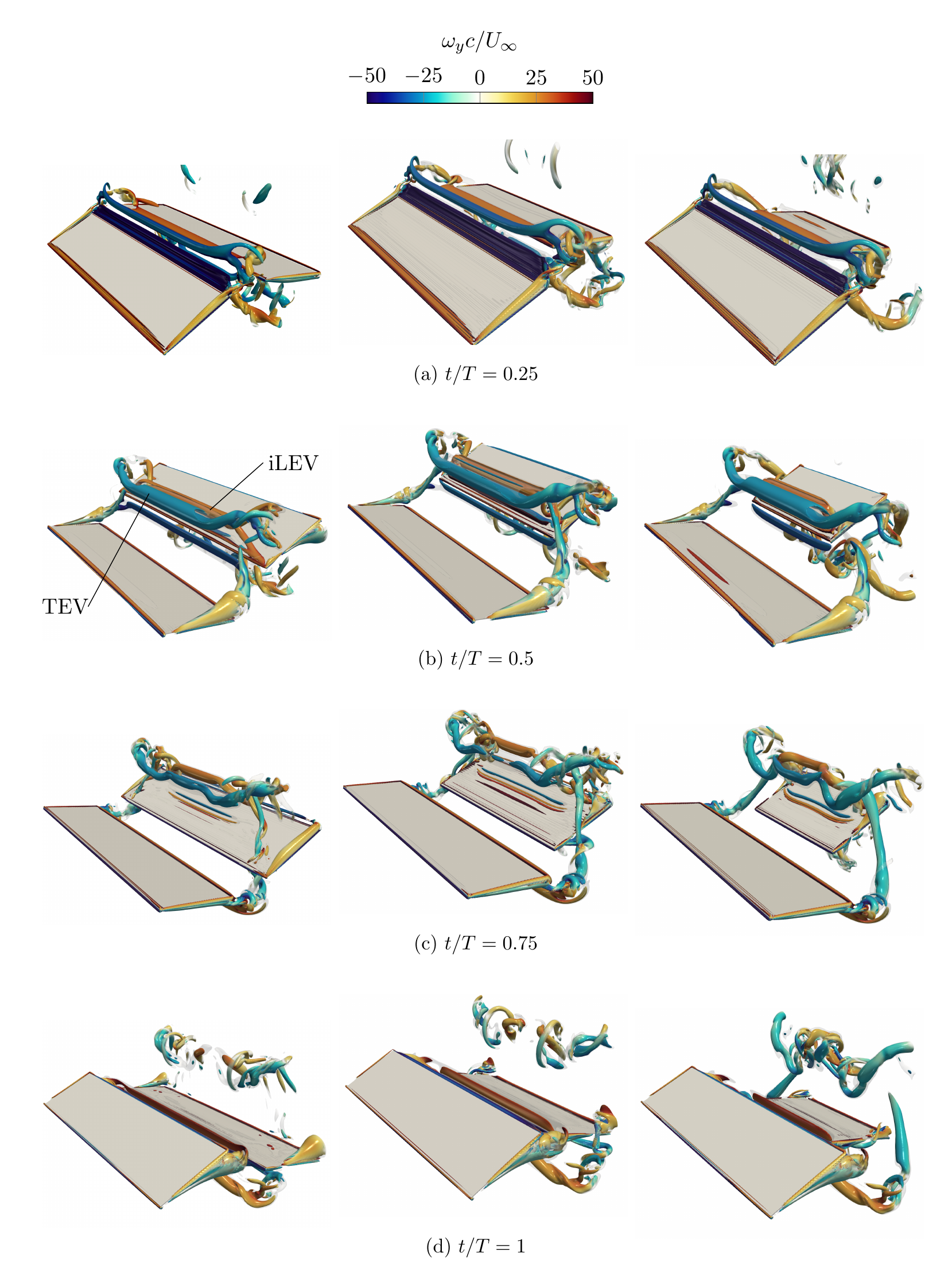}
   \caption{Flow visualization of vortical structures for cases AR44, AR43 and AR42 (from left to right). 
    Times correspond to the hindwing's (a) mid upstroke, (b) end upstroke, (c) mid downstroke, and (d) end downstroke. }
  % \label{fig:fig1}
   \label{fig:3Dvort_a}
\end{figure}

\begin{figure}
   \centering
	\includegraphics[width=1\textwidth]{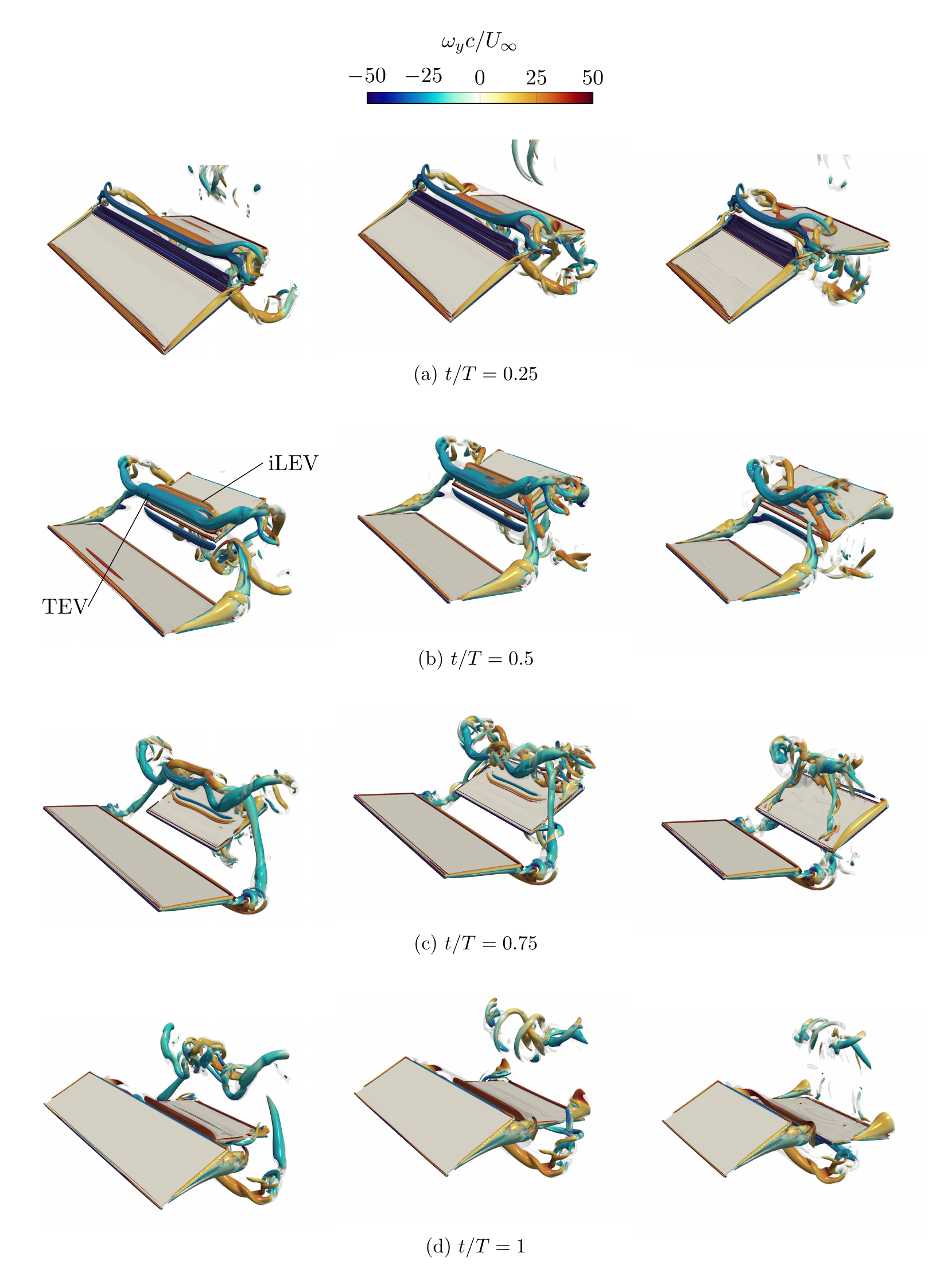}
    \caption{Flow visualization of vortical structures for cases AR42, AR32 and AR22 (from left to right).
    Times correspond to the hindwing's (a) mid upstroke, (b) end upstroke, (c) mid downstroke, and (d) end downstroke. 
    }
	\label{fig:3Dvort_b}
\end{figure}

We start analyzing qualitatively the interactions between the hind- and the forewings using the flow visualizations presented in Figures \ref{fig:3Dvort_a} and \ref{fig:3Dvort_b}.
Isosurfaces of the second invariant of the velocity gradient tensor, $Q = 6\Omega^2_0$, are used to represent the vortical structures around the wings \citep{Hunt1988}, where $\Omega_0 = 2 \pi f$ is the angular velocity of oscillatory motion of the wings. 
These isosurfaces are colored with the spanwise vorticity, $\omega_y$, to visualize the direction and intensity of the rotation of the vortices.
Note that, since the focus is on the interactions between both wings, the wake region (i.e., $x>1.25c$) has been clipped in Figures \ref{fig:3Dvort_a} and \ref{fig:3Dvort_b}. 
The figures show four different snapshots of the vortical structures during the forewing's upstroke (which corresponds to the hindwing's downstroke),  
providing an overview of the time evolution of the structures for all simulated cases. Animations are also provided in the supplementary material to show the interaction of the vortical structures during the complete oscillation cycle.

The overall evolution of the vortical structures that drive the interaction between the fore- and hindwing is as follows. 
The forewing's downstroke generates a trailing edge vortex (TEV, colored in blue in Figures \ref{fig:3Dvort_a} and \ref{fig:3Dvort_b}), which induces a counter-rotating leading edge vortex (iLEV, colored in orange) as it approaches the leading edge of the hindwing. 
These two vortices form a dipole (Figures \ref{fig:3Dvort_a}b and \ref{fig:3Dvort_b}b), which moves downstream over the upper surface of the hindwing. 
Eventually, the dipole becomes unstable (Figures \ref{fig:3Dvort_a}c and \ref{fig:3Dvort_b}c), which leads to the vortex breakdown (Figures \ref{fig:3Dvort_a}d and \ref{fig:3Dvort_b}d). Note that the structures showed by Figures \ref{fig:3Dvort_a}a and \ref{fig:3Dvort_b}a are equivalent to those of the downside of Figures \ref{fig:3Dvort_a}c and \ref{fig:3Dvort_b}c, respectively, with opposite sign for vorticity. 
The figures and the animations show that the breakdown process of the dipole seems to start near the wing tips, progressing towards the midspan ($y=0$) of the hindwing as the dipole is advected downstream.  

Figure \ref{fig:3Dvort_a}b shows that there is a region at the midspan where both TEV and iLEV are uniform in the spanwise direction (i.e, they are quasi-2D structures), as previously discussed in our previous work \citep{Arranz2020}. 
This spanwise uniformity is lost near the wing tips due to the bending of the iLEV and TEV, and the interaction with the wing tip vortices.
Indeed, near the wing tips, the TEV and the iLEV merge with the tip vortices of the hindwing and the forewing, forming a Y-branching bifurcation clearly visible in Figure \ref{fig:3Dvort_a}b. 
The interactions among the iLEV, the TEV and the tip vortices depend on the aspect ratio of both wings. 
Therefore, it is interesting to evaluate the extension of the various
regions for each of the cases, also in view of their possible impact on the load distribution. 
In order to analyze these effects, we are going to consider two groups of cases.
The first group (AR4X, plotted in Figure \ref{fig:3Dvort_a}) contains cases with constant aspect ratio of the forewing ($\AR_f=4$) and variable aspect ratio of the hindwing ($\AR_h = 4, 3$ and 2). 
The second group (ARX2, plotted in Figure \ref{fig:3Dvort_b}) contains cases with variable aspect ratio of the forewing ($\AR_f=4, 3$ and 2) and constant aspect ratio of the hindwing ($\AR_h=2$).

From the point of view of the TEV, the analysis of AR4X (Figure \ref{fig:3Dvort_a}) and ARX2 (Figure \ref{fig:3Dvort_b}) suggests that the geometry and the breakdown of the TEV is governed by the aspect ratio of the forewing, $\AR_f$: 
Figure \ref{fig:3Dvort_a} shows that the geometry of the TEV at time $t=0.5T$ and $t=0.75T$ is largely independent of $\AR_h$. 
On the other hand, Figure \ref{fig:3Dvort_b} shows clear changes in the geometry and evolution of the TEV as the aspect ratio of the forewing decreases.
In particular, the region where the TEV is uniform in the spanwise direction is reduced as $\AR_f$ becomes smaller. 
Also, the breakdown of the TEV  occurs earlier as $\AR_f$ becomes smaller (compare cases ARX2 in Figure \ref{fig:3Dvort_b}), consistent with the idea that the breakdown originates at the interaction of the TEV with the wing tip vortices of the forewing.
Indeed, Figures \ref{fig:3Dvort_a}c and \ref{fig:3Dvort_b}c clearly show that the TEV is connected to the wing tip vortices formed on the forewing during the upstroke, although that connection is interrupted when $\AR_h$ becomes comparable to $\AR_f$.

From the point of view of the iLEV, Figures \ref{fig:3Dvort_a} and \ref{fig:3Dvort_b} show that this vortex is quasi-2D whenever the quasi-2D region of the TEV is larger than the span of the hindwing. 
This implies that the structure of the iLEV not only depends on the size of the hindwing but also on the size of the forewing.
Figures \ref{fig:3Dvort_a} and \ref{fig:3Dvort_b} also show that the iLEV seems to be connected to the wing tip vortices generated on the downstroke of the hindwing, 
and that these wing tip vortices appear to be weaker when  $\AR_f > \AR_h$ than when $\AR_f = \AR_h$. 
On the other hand, Figures \ref{fig:3Dvort_a} and \ref{fig:3Dvort_b} show that the wing tip vortices of the forewing do not vary with $\AR_f$ or $\AR_h$, being the same for all cases in AR4X and ARX2. 
This observation is consistent with the results reported for heaving and flapping wings at comparable Reynolds numbers \citep{Gonzalo2018}.

%%%%%%%%%%%%%%%%%%%%%%%%%%%%%%%%%%%%%%%%%%%%%%%%%%%%%%%%%%%%%%%%%%%%%%%%%%%%%%%
%% FIGURE 3D VORTICES
%%%%%%%%%%%%%%%%%%%%%%%%%%%%%%%%%%%%%%%%%%%%%%%%%%%%%%%%%%%%%%%%%%%%%%%%%%%%%%%

%%%%%%%%%%%%%%%%%%%%%%%%%%%%%%%%%%%%%%%%%%%%%%%%%%%%%%%%%%%%%%%%%%%%%%%%%%%%%%%
%% FIGURE 3D VORTICES
%%%%%%%%%%%%%%%%%%%%%%%%%%%%%%%%%%%%%%%%%%%%%%%%%%%%%%%%%%%%%%%%%%%%%%%%%%%%%%%

\subsection{Aerodynamic forces} \label{sec:forces}

In \S \ref{sec:flow_struct} we have seen that
the aspect ratio of both wings influences the structure of the vortices and their interactions. 
We turn now our attention to the aerodynamic performance of each wing,
by analyzing independently the forces produced by the fore- and the hindwing.
%

%%%%%%%%%%%%%%%%%%%%%%%%%%%%%%%%%%%%%%%%%%%%%%%%%%%%%%%%%%%%%%%%%%%%%%%%%%%%%%%
% FIGURE TOTAL FORCE
\begin{figure}
	\centering
    \includegraphics[width=\textwidth]{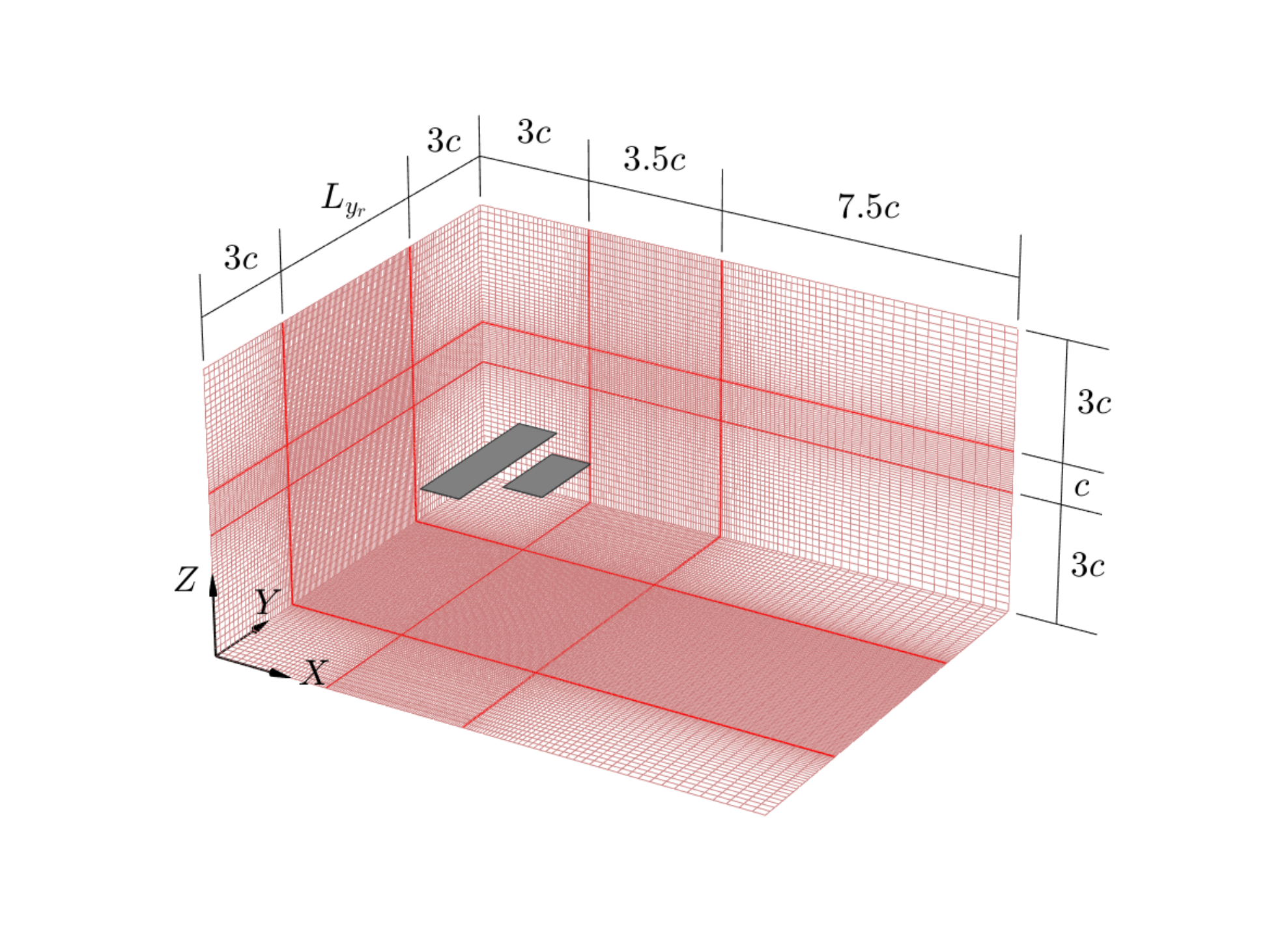}
    \caption{Total horizontal and vertical force coefficients of the forewing as a function of time (shaded area denotes the downstroke). AR44 \lyy{-}{matlabdef1}, AR43 \lyy{-}{matlabdef2}, AR42 \lyy{-}{matlabdef3}, A32 \lyy{-}{matlabdef4}, AR22 \lyy{-}{matlabdef5}, AR4 \lyy{--}{matlabdef3} and AR2 \lyy{--}{matlabdef5}.
    }
    \label{fig:totalForce2}
\end{figure}
%%%%%%%%%%%%%%%%%%%%%%%%%%%%%%%%%%%%%%%%%%%%%%%%%%%%%%%%%%%%%%%%%%%%%%%%%%%%%%%

We start analyzing the total force produced by the forewing, by looking
at time histories of force coefficients in Figure \ref{fig:totalForce2} 
and their time-averaged values in Table \ref{tab:MeanForce}.
Note that for the forewing the incoming flow is unperturbed resulting in a simpler aerodynamic response than for the hindwing, that is subjected to the wake of the forewing.
Figure \ref{fig:totalForce2} shows the time evolution of both the horizontal, $C_{x,f} (t)$, and the vertical, $C_{z,f} (t)$, force coefficients of the forewing for all cases during a cycle.
The shaded area corresponds to the downstroke motion of the forewing ($0 \leq t/T \leq 0.5$).
The time history of the aerodynamic force coefficients is similar to that obtained
for isolated wings in heaving and pitching motion \citep{dong2006wake}.
The thrust (negative $C_{x,f} (t)$) presents two peaks in the middle of the downstroke ($t/T = 0.25$) and the upstroke ($t/T = 0.75$), namely when the wing heaves at maximum velocity and the pitch angle projects the normal force forward. 
Thus, thrust is produced both during the down- and the upstroke motions.
The lift ($C_{z,f} (t)$) presents alternating positive/negative peaks at these instants, resulting in a vanishing mean lift due to the symmetry of the motion.

Although all cases show a similar time history of the force coefficients, there are differences in the intensity of the peak forces.
The comparison of cases AR4X shows that the peak values increase as the $\AR_h$ decreases (see inset), suggesting that  increasing the aspect ratio of the hindwing results in a progressive reduction of the aerodynamic load of the forewing.
However, these differences are  small, so that they have a limited impact
in the resulting time-averaged force coefficients. 
In particular, 
the time averaged values show that the thrust coefficient increases about $2\%$ when the hindwing aspect ratio is reduced from $\AR_h = 4$ to $\AR_h = 3$ but no further increase is observed if the $\AR_h$ is further decreased or even in the limiting case in which the hindwing is not present (case AR4).
Note that for the forewing, the cases AR4X show the influence of the presence
of hindwings of different sizes in the performance of a forewing of fixed
aspect ratio. 
On the other hand the cases ARX2 mainly show the effect of varying the aspect
ratio of the forewing itself, as evidenced by the comparison with the
corresponding cases of isolated wing.
Thus, the time-averaged thrust coefficient of the case AR2 is the same as the one of case AR22, and the same is true for cases AR4 and AR42 (see Table \ref{tab:MeanForce}).
For the lift coefficient, a similar observation can be made 
although the values differ slightly.

%%%%%%%%%%%%%%%%%%%%%%%%%%%%%%%%%%%%%%%%%%%%%%%%%%%%%%%%%%%%%%%%%%%%%%%%%%%%%%%%%%%%%%%%%%%%%%%%%%%%%%%%%%%%%%%%%%%%%%%%
\begin{table}
	\begin{center}
		\begin{tabular}{ccccccc}

			  Case  & $\overline{C}_{T,f}$ & $\overline{C}_{T,h}$ & $\overline{C}_{z,f}$ & $\overline{C}_{z,h}$ & $\eta_f$ & $\eta_h$  \\
			\hline
			AR44  &  0.91  &  0.65  &  4.39  &  2.24  &  0.23  &  0.20     \\
			AR43  &  0.93  &  0.63  &  4.47  &  1.96  &  0.23  &  0.20     \\
			AR42  &  0.93  &  0.62  &  4.52  &  2.01  &  0.23  &  0.19     \\
			AR32  &  0.89  &  0.59  &  4.24  &  1.84  &  0.23  &  0.20     \\
			AR22  &  0.81  &  0.57  &  3.83  &  2.05  &  0.23  &  0.21   \\
            \hline
			AR4   &  0.93  &    -    &  4.58  &    -    &  0.22  &    -       \\
			AR2   &  0.81  &    -    &  3.95  &    -    &  0.22  &    -       

		\end{tabular}
	\end{center}
			\caption{Force coefficients averaged on time and propulsive efficiency for each wing and case. Thrust coefficient is averaged over a full cycle and lift coefficient is averaged over a half cycle.}
			\label{tab:MeanForce}
\end{table}

%%%%%%%%%%%%%%%%%%%%%%%%%%%%%%%%%%%%%%%%%%%%%%%%%%%%%%%%%%%%%%%%%%%%%%%%%%%%%%%
% FIGURE TOTAL FORCE
\begin{figure}
	\centering
    \includegraphics[width=\textwidth]{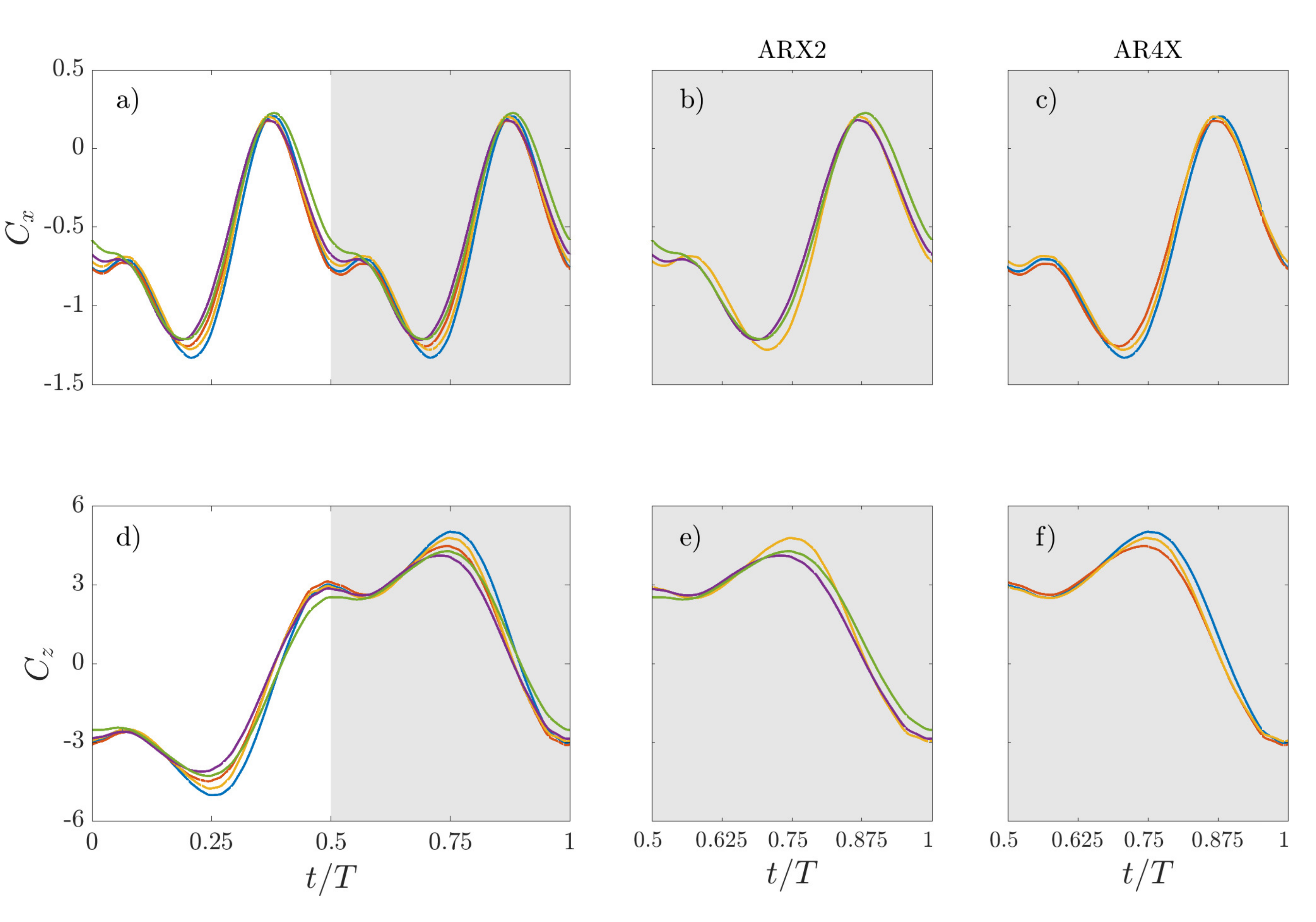}
    \caption{
    Time evolution of the (a-c) horizontal and (d-f) vertical force coefficients of the hindwings during its downstroke. All cases are shown in (a,d). Selected cases are shown in (b,c,e,d). 
    AR44 \lyy{-}{matlabdef1}, AR43 \lyy{-}{matlabdef2}, AR42 \lyy{-}{matlabdef3}, AR42 \lyy{-}{matlabdef3}, A32 \lyy{-}{matlabdef4} and AR22 \lyy{-}{matlabdef5}. 
    }
    \label{fig:hindTforceHalf}
\end{figure}
%%%%%%%%%%%%%%%%%%%%%%%%%%%%%%%%%%%%%%%%%%%%%%%%%%%%%%%%%%%%%%%%%%%%%%%%%%%%%%%

To analyse the effect of varying $\AR_f$ and $\AR_h$ on the aerodynamic performance of the hindwing, Figures \ref{fig:hindTforceHalf}a 
and \ref{fig:hindTforceHalf}d show the total horizontal and vertical force coefficients, 
respectively, for the hindwing of all cases.
In addition, the downstroke of the cases ARX2 and AR4X is presented separately in 
Figures \ref{fig:hindTforceHalf}b and \ref{fig:hindTforceHalf}e (ARX2) and Figures \ref{fig:hindTforceHalf}c and \ref{fig:hindTforceHalf}f (AR4X).
Figure \ref{fig:hindTforceHalf} clearly shows that the vortical interactions of the TEV and the iLEV have an impact on the aerodynamic performance.
The main difference is related to  the magnitude of the aerodynamic force coefficients, which are  considerably smaller for the hindwing than for the forewing.
Similarly to the forewing, the peak of both forces occurs at $t/T \approx 0.25$ for the upstroke and $t/T \approx 0.75$ for the downstroke.
However, additional peaks appear in $C_{x,h}$ and $C_{z,h}$ at the stroke reversals, corresponding to the formation of the iLEV on the suction surface of the hindwing. 

As for the forewing discussed above, the time evolution of the force produced by the hindwing shows little variation with $\AR_f$ and $\AR_h$.
The differences among the cases occur mainly at the time instants where the heaving velocity is maximum ($t/T=0.25$ and $0.75$ ) and at the stroke reversals ($t/T=0.5$ and $1$).
The analysis of the time histories in Figure \ref{fig:hindTforceHalf} and the time-averaged values in Table \ref{tab:MeanForce} shows that for the hindwing is more difficult to identify the trends, especially for the lift coefficient.
The time-averaged thrust coefficient of the hindwing decreases monotonically when reducing its  aspect ratio (for cases AR4X) and
also when decreasing the aspect ratio of the forewing (for cases ARX2).
On the contrary, the trend in the time-averaged lift coefficient is not monotonous. 
For example, for cases AR4X it is observed a 
maximum value of $\overline{C}_{z,h} = 2.24$ for case AR44, which then decreases for case AR43 to a value of $\overline{C}_{z,h} = 1.96$ and a further
decrease of the aspect ratio of the hindwing leads to an increase to a value of $\overline{C}_{z,h} = 2.01$ for case AR42.
A similar observation can be made for cases ARX2. 
This non-monotonic trend might be related to the fact that the force coefficients are strongly influenced by the 
structure of the iLEV. 
As shown before, the iLEV depends both on the size of forewing and hindwings, in a non-trivial way.
%

%%%%%%%%%%%%%%%%%%%%%%%%%%%%%%%%%%%%%%%%%%%%%%%%%%%%%%%%%%%%%%%%%%%%%%%%%%%%%%%
%% FIGURE COLORPLOT
\begin{figure}
   \centering
	\includegraphics[width=1\textwidth]{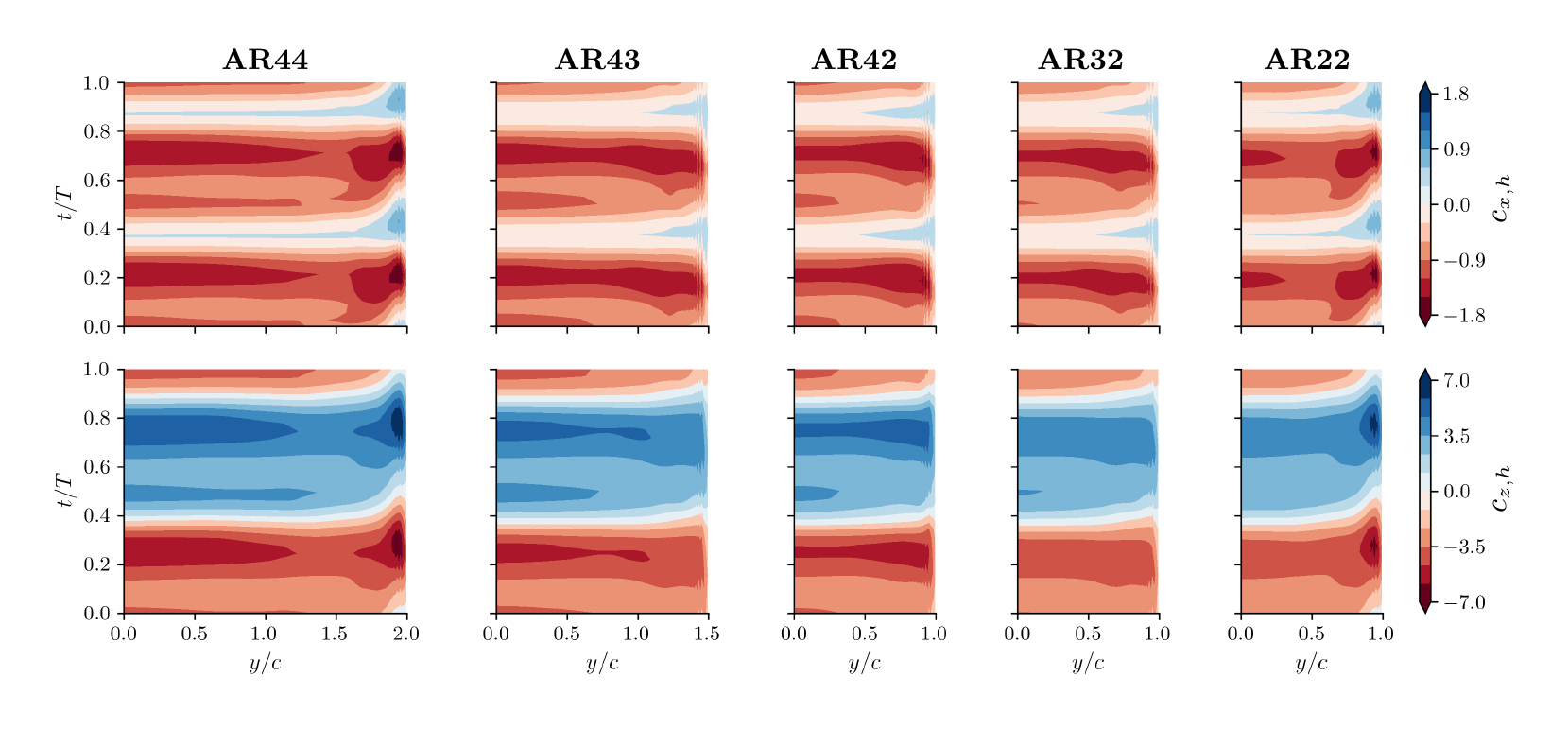}
    \caption{Horizontal, $c_{x,h}$ (top row), and vertical, $c_{z,h}$ (bottom row), force coefficients of the hindwing as a function of span and time.}
	\label{fig:Stcoeff}
\end{figure}

The previous results show that it is not an easy task to relate the vortical structures to the global force coefficients, 
and to look for further insight it is worth to analyse the sectional force coefficients.
Figure \ref{fig:Stcoeff} shows a colormap of the horizontal, $c_{x,h}$, and vertical, $c_{z,h}$, sectional force coefficients of the hindwing as a function of the span and time.
Since the force distribution is symmetric with respect to the mid-span, we only show half of the span, from mid-span to the wing tip.
In accordance to what was observed when discussing the total force coefficients, Figure \ref{fig:Stcoeff} shows that the hindwing mostly produces thrust (i.e., negative $c_{x,h}$, in red), both during the upstroke ($0<t/T<0.5$) and the downstroke ($0.5<t/T<1$) motions, whereas the lift, $c_{z,h}$, shows opposite sign during the upstroke and the downstroke.

Looking first at the sectional force coefficient of the cases ARX2, 
 we observe that as the $\AR_f$ decreases the spanwise loading becomes less uniform and the maximum intensity of the forces (found at $t/T = 0.25$ and $0.75$) also decreases. 
 This is not obvious when comparing cases AR42 and AR32, but it is very clear when comparing any of these two cases to the case AR22.
 This observation can be directly connected to the structure of the iLEV shown in Figure \ref{fig:3Dvort_b}b, that
 shows a quasi-2D iLEV for cases AR42 and AR32 while the iLEV of case AR22 is rather 3D and strongly influenced by the
 wing tip vortices.
 Regarding the sectional force coefficient of the cases AR4X, decreasing the aspect ratio of the hindwing
 leads to a more uniform spanwise loading. 
 This can also be linked to the vortical structures in Figure \ref{fig:3Dvort_a}b.
 For case AR44, there is a central region where the iLEV is quasi-2D, but there is significant area near the
 tips influenced by the three-dimensionality of the wing tip vortices. 
 Accordingly, for this case the sectional force coefficient display force peaks near the wing tips.
 Both the force peaks near the wing tips and the area of vortical three-dimensionality tend to disappear when decreasing the aspect ratio of the hindwing. 
 
 Note also that near the wing tips, the spanwise loading  of case AR44 just discussed is rather similar in shape and size to the spanwise loading of case AR22, including the aforementioned force peaks although with a somewhat different intensity. 
 This suggests that when both wings have the same aspect ratio (and as consequence the wing tips of the fore- and the hindwing are aligned in the streamwise direction), 
 the mechanisms of interaction of the 
 vortical structures near the wing tips are very similar. 
 Indeed, the Y-branching flow structures observed at the wing tips of the hindwing 
 are comparable in both cases (compare first column of Figure \ref{fig:3Dvort_a} and third column of Figure \ref{fig:3Dvort_b}).
 A further observation is that the force peak at the hindwing tips presents a time lag with respect to the sectional force coefficient at the mid-span.
This time lag can also be observed for the sectional force coefficients of the forewings (not shown here) and in single wings in heaving motion \citep{Gonzalo2018,visbal2013three}, which suggests that this time lag is not a consequence of the TEV-iLEV interaction.

Finally, the sectional force coefficients also shed some light into the non-monotonic trend 
of the time-averaged lift coefficient  discussed above.
We first look at the cases AR4X. Comparing the case AR43 with the case AR42, 
we see that for AR43 the intensity of the lift coefficient at $t/T=0.25$ (i.e., near the maximum) decreases towards the tip, 
while this does not happen for AR42.
Consequently, the time averaged lift coefficient is smaller for case AR43.
Comparing cases AR44 and AR43, it can be observed that the lift coefficient of both cases is rather similar but the contribution of the wing tip force peak in case AR44 makes this case to exhibit a larger time averaged lift coefficient.
For the cases ARX2, the explanation of the non-monotonic trend is somewhat different. 
When decreasing the aspect ratio of the forewing, the intensity of the TEV is reduced, resulting in a lower level 
of lift coefficient overall. 
So that the lift coefficient of case AR32 is smaller than the one of case AR42 over the whole wing. However, 
when further decreasing the aspect ratio of the forewing to $\AR_f=2$, 
the discussed-above force peak at the tip appears leading
to a slightly larger value of the time-averaeged lift coefficient of the case AR22 compared to the case AR32.

\subsection{Power requirements and propulsive efficiency}

\begin{figure}[h!]
	\centering
    \includegraphics[width=\textwidth]{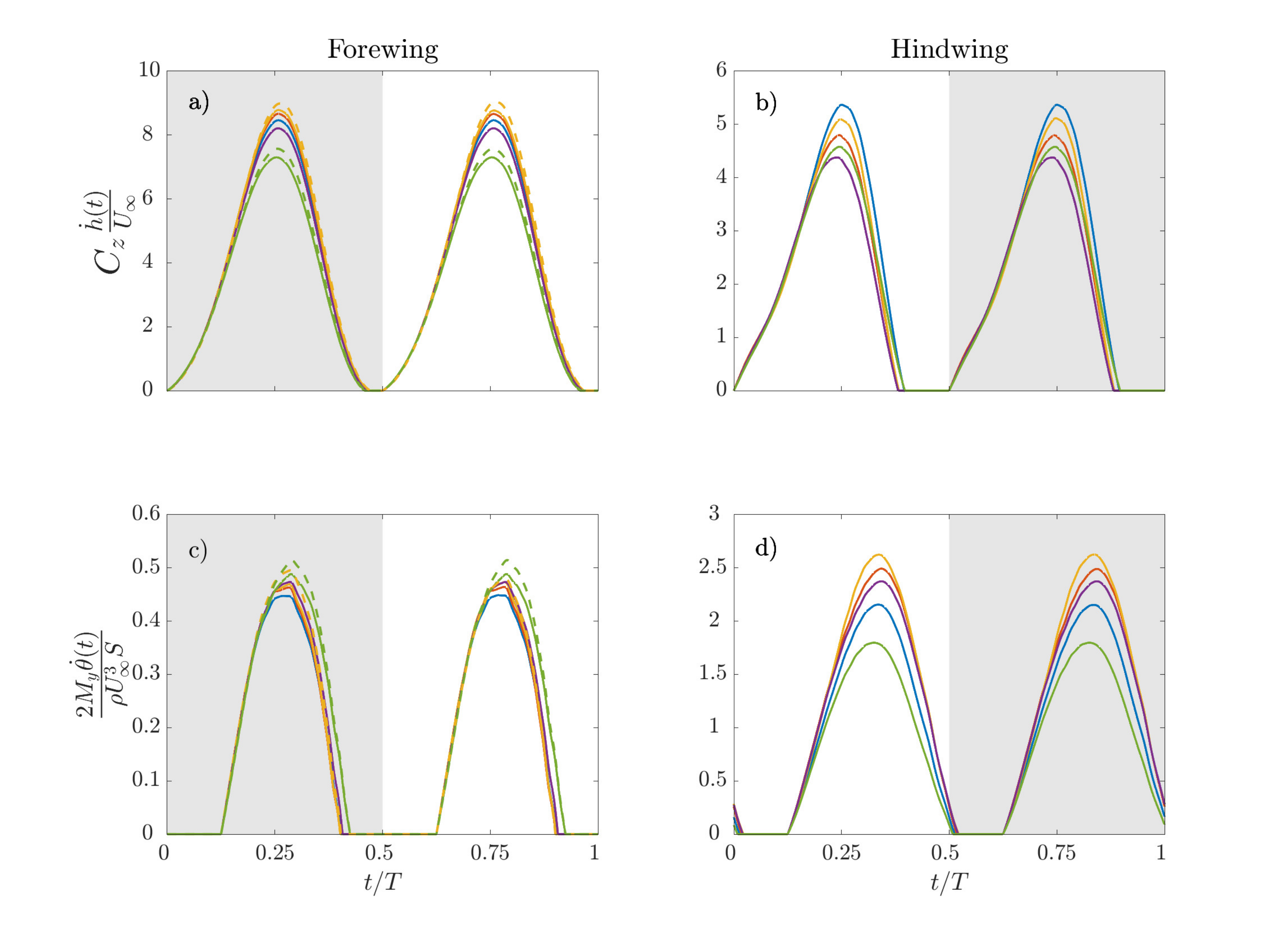}

    \caption{(a,b) Heaving power input. (c,d) Pitching power input. Shaded region  indicates the downstroke. 
    AR44 \lyy{-}{matlabdef1}, AR43 \lyy{-}{matlabdef2}, AR42 \lyy{-}{matlabdef3}, A32 \lyy{-}{matlabdef4} and AR22 \lyy{-}{matlabdef5}, AR4 \lyy{--}{matlabdef3} and AR2 \lyy{--}{matlabdef5}.}
     \label{fig:powerInput}
\end{figure}

After discussing the vortical structures and the aerodynamic forces, we conclude the article with a discussion
of power requirements. 
According to the literature, the main benefit of flapping wings in horizontal tandem over single flapping wings is associated to an increase in the propulsive efficiency of the hindwing \citep{akhtar2007hydrodynamics,broering2013investigation,broering2015numerical}. 
For the analysed cases in this work a subtle improvement of $\sim 4\%$ is produced upon the forewing when compared to an isolated wing. 
However, in the case of the hindwing the increase in the time averaged thrust coefficient due to the quasi-2D vortical interaction does not translate into an improved propulsive efficiency of the hindwing,
as can be seen in the efficiencies reported in Table \ref{tab:MeanForce}, in particular for cases ARX2. 
For instance, 
case AR42 generates $\sim 9 \%$ more net thrust and $\sim 2 \%$ less net lift than case AR22 but 
the propulsive efficiency of case AR42 is $\sim 9 \%$ lower than the propulsive efficiency of case AR22.
A similar behaviour is observed when comparing cases AR44 and AR42: a small (i.e. $\sim 5\%$) decrease in $\overline{C}_{T,h}$ as $\AR_h$ decreases, together with a larger (i.e. $\sim 10\%$) decrease in the vertical force coefficient results in a lower propulsive efficiency for the hindwing in AR42 than in AR44. 
Note that in principle less lift is beneficial from the point of view of power requirement to move the wing.
Clearly, in the present case, looking only at the values of thrust and lift is not sufficient, and it is necessary to analyze
the power requirements due to the pitching motion (second term in Eq. \eqref{eq:powerinput}).

Figure \ref{fig:powerInput} shows the power input associated independently to the heaving and pitching motions, first and second term in Eq. \eqref{eq:powerinput}, respectively, for each wing and case.
First, it is interesting to note that even if the overall power requirements of hindwing and forewing are similar (i.e., $\overline{P}_h \sim \overline{P}_f$), the pitching motion represents less than $5\%$ of the power requirement of the forewings, but about $35\%$ of the power requirement of the hindwings (note the different scale in the ordinate axis in the panels of Figure \ref{fig:powerInput}). 
The non-negligible contribution of pitching moment to the power requirements of the hindwing is a consequence of the suction force generated by the traveling pair of vortical structures (TEV-iLEV) over the hindwing.
This moves the center of pressure of the wing backwards,
and explains why the variations in $\overline{C}_z$ and $\overline{C}_T$ cannot be used to predict the variations in $\eta_h$. 
In addition, it is worth noting that the power requirements associated to the heaving motion of both wings peak at mid-stroke, when the aerodynamic forces and heaving velocity are maxima (see Figures \ref{fig:powerInput}a and b).
However, the power requirements associated with the pitching motion tend to present a lag behind the maximum heaving velocity.
This effect is more noticeable for the hindwing (Figure \ref{fig:powerInput}d), with maximum power occurring close to the time of maximum pitching velocity (i.e., $t/T \approx 0.375$ and 0.875). 

In terms of the relation of the two-dimensionality of the flow observed by the hindwing and the power required to sustain the motion, Figure \ref{fig:powerInput}d shows that the power requirements associated to the pitching motion of the hindwing increase when the flow is made more 2D over the hindwing.
This occurs both when the $\AR$ of the forewing is increased (i.e. cases ARX2), and when the $\AR$ of the hindwing is reduced (i.e. cases AR4X). 
On the other hand, the power required for the heaving motion decreases when $AR_h$ is reduced (i.e. compare cases AR4X in Figure \ref{fig:powerInput}b) and also when the $\AR_f$ is increased (i.e. case ARX2), although this reduction is not monotonic with the reduction of the corresponding wing as mentioned before. 
Overall, a more two-dimensional interaction of the forewing's wake with the hindwing, either by reducing the $\AR_h$ or by increasing the $\AR_f$, leads to a larger contribution of the pitching motion to the power required to sustain the motion of the wing, which is translated to a small reduction of the efficiency of the hindwing (see Table \ref{tab:MeanForce}).

\section{Conclusions}
\label{conclusions}

We have presented direct numerical simulations of the flow around two flapping wings in horizontal tandem configuration. 
The simulations were performed at $Re=1000$ for wings undergoing a heaving and pitching motion, 
at Strouhal number $St_c = 0.7$, that corresponds to optimal 2D kinematics \citep{Ortega2019}.
For this configuration, the aerodynamic performance of the system is dictated by the interactions between 
the trailing edge vortex (TEV) shed from the forewing and the induced leading edge vortex (iLEV) formed on the hindwing. 
This interaction is also modified by the aspect ratio of the wings, 
resulting in quasi-2D regions for sufficiently large aspect ratios (i.e. at least $\AR = 4$).  
The aim of this study was to analyse the aerodynamic performance of the two wings system focusing on the quasi-2D interaction between TEV and iLEV and how this translates into the spanwise load distribution on the hindwing.
Several tandem and isolated configurations were simulated, varying the aspect ratio of both fore- and hindwings. 
With the results of these simulations we have presented here a qualitative  characterization of the vortical structures, and a quantitative analysis of the load distribution and aerodynamic performance of both wings. 

The aerodynamic performance of the forewing is similar to that of an isolated wing, with a small modulating effect produced by the forewing-hindwing interactions. 
The thrust coefficient of the forewing is slightly increased when 
the hindwing becomes smaller than the forewing (i.e.,  $\AR_f > \AR_h$) with respect to case in which $\AR_f = \AR_h$.
Interestingly, no significant differences were observed in the thrust coefficient between cases with $\AR_f > \AR_h$ at constant $\AR_f$.  
In contrast, the lift coefficient increases monotonically when decreasing the $\AR_h$, although this variation is limited ($3\%$ increase).
Regarding the propulsive efficiency of the forewing, the results show that it is slightly increased ($\eta_f = 0.23$) with respect to an isolated wing ($\eta = 0.22$).

On the other hand, the aerodynamic performance of the hindwing is clearly different to that of an isolated wing, as a consequence of the interaction with the forewing's TEV.
The spanwise load distribution is determined by the iLEV, whose structure depends on the spanwise length of the TEV. 
The latter is dictated by the aspect ratio of the forewing only. 
For instance, when $\AR_f > \AR_h$ the TEV promotes a quasi-2D iLEV over the whole span of the hindwing, yielding an almost uniform spanwise load distribution.
At a constant $\AR_h = 2$, a more uniform load distribution results in larger thrust coefficients on the hindwing (increasing a 8\% when $\AR_f$ increases from 2 to 4), with a non-monotonic variation of the lift coefficient. 
The latter is related to the role of the wing-tip vortices of the hindwing, and their contribution to the aerodynamic forces.  

Overall, these vortex interactions result in lower aerodynamic loads on the hindwing when compared to the forewing and/or isolated wings for all the configurations analyzed here. 
However, the total power requirements for the motion of the hindwing remains similar to those of the forewing, resulting in a lower propulsive efficiency for the hindwing ($\eta_h \sim 0.20$) compared to the forewing. 
Note that the total power requirement for the motion of the wings is split into a contribution from the heaving motion, and a contribution from the pitching motion. 
For the isolated wing and for the forewing, the contribution from the pitching motion is small, and around $95\%$ of the total power comes from the heaving motion. 
For the hindwing, the power requirements associated to the heaving motion are drastically reduced compared to the forewing. 
However, the contribution from the pitching motion is increased by a factor of $5$, resulting in a 65/35\% split between the power required between the heaving and pitching motions.

%%%%%%%%%%%%%%%%%%%%%%%%%%%%%%%%%%%%%%%%%%%%%%%%%%%%%%%%%%%%%%%%%%%%%%%%%%%%%%%%%%%%%
\section*{Acknowledgements}
%%%%%%%%%%%%%%%%%%%%%%%%%%%%%%%%%%%%%%%%%%%%%%%%%%%%%%%%%%%%%%%%%%%%%%%%%%%%%%%%%%%%%

This work was partially supported by the State Research Agency of Spain (AEI) under grant DPI2016-76151-C2-2-R including funding from the European Regional Development Fund (ERDF).
The computations were performed at the supercomputer Tirant from the {\it Red Espa\~nola
de Supercomputaci\'on} in activity IM-2019-3-0011. 

\section*{Data Availability Statement}

The data that support the findings of this study are available from the corresponding author upon reasonable request.

\bibliographystyle{unsrt}
\bibliography{paper}

\end{document}